\documentclass[12pt,twoside]{article}

\usepackage{fleqn,espcrc1}

\usepackage{graphicx}
\usepackage[figuresright]{rotating}

\usepackage{psfrag}
\usepackage{epsfig}
\usepackage{psfig}
\def\3{{\ss}}
\newcommand{\beq}{\begin{equation}}
\newcommand{\eeq}{\end{equation}}
\newcommand{\beqa}{\begin{eqnarray}}
\newcommand{\eeqa}{\end{eqnarray}}
\newcommand{\no}{\nonumber }

\newcommand{\AmS}{{\protect\the\textfont2
  A\kern-.1667em\lower.5ex\hbox{M}\kern-.125emS}}

\title{Chiral Dynamics in Few--Nucleon Systems}

\author{E.~Epelbaum\address{Forschungszentrum J\"ulich, Institut f\"ur 
                 Kernphysik (Theorie), \\
                 52425 J\"ulich, Germany},  %
        H.~Kamada$^{\rm{a},}$\address{Institut f\"ur  Theoretische Physik II,
         Ruhr-Universit\"at Bochum, \\
        52425 J\"ulich, Germany},
        A. Nogga$^{\rm{b}}$,
        H. Wita\l a\address{Institute of Physics, Jagellonian University, \\ 
                 PL-30059 Cracow, Poland}, 
        W.~Gl\"ockle$^{\rm{b}}$,
        and
        Ulf-G.~Mei\3ner$^{\rm{a}}$}
       
\begin{document}
\input amssym.def
\input amssym

\maketitle

\begin{abstract}
We employ the chiral nucleon--nucleon potential derived using the method
of unitary transformation up to next--to--next--to--leading order (NNLO)
to study bound and scattering states in the two--nucleon system. 
The predicted partial wave phase shifts and mixing
parameters for higher energies and higher
angular momenta beyond the ones which are fitted are mostly well described for energies below 300~MeV.
Various deuteron properties are discussed.
We also apply the next--to--leading order (NLO) potential to 3N and 4N systems.  The resulting 3N and 4N binding energies are
in the same range what is found using standard NN potentials. Experimental low--energy
3N scattering observables are also very well reproduced like for standard NN forces. Surprisingly
the long standing $A_y$--puzzle is resolved at NLO. The cut-off
dependence of the scattering observables is rather mild.
\end{abstract}

\section{INTRODUCTION}

Over the past decade effective field theory methods have been successfully applied 
to many different physical processes and became, in particular, a standard tool in calculating low--energy 
properties of the purely pionic and pion--nucleon systems. In this method one starts from the most 
general Lagrangian for pions and nucleons. Apart from the usual symmetry constraints, spontaneously and explicitly broken 
chiral symmetry of QCD is taken into account. As a consequence, the  couplings of pion fields,
which play the role of Nambu-Goldstone bosons within this formalism, 
are of derivative type.
As has been shown by Weinberg more than 20 years ago, one can use such an effective Lagrangian to derive the 
expansion of low energy S-matrix elements in powers of some 
scale $Q$ related to low external momenta of pions and nucleons \cite{We79} 
(the 4-momenta of the external pions and the 3-momenta
of the external nucleons). 

Motivated by successful applications of CHPT in the 
$\pi \pi$ and $\pi N$ sectors, Weinberg proposed in 1990
to extend the formalism to the NN interaction \cite{We90}.
The crucial difference is, however, that the nucleon--nucleon 
interaction is non--perturbative at low energies: it is strong enough
to bind two nucleons in the deuteron.  
Thus, direct application of CHPT to the NN amplitude will
necessarily fail. The way out of this problem, proposed by 
Weinberg, is to apply CHPT not to the amplitude but to a 
kernel of the corresponding integral Lippmann--Schwinger (LS) equation
using time--ordered perturbation theory. Such a kernel or effective NN potential is then defined 
as a sum of all diagrams with two incoming
and outgoing nucleons that do not contain purely nucleonic intermediate states (reducible diagrams). 
A systematic power counting for the potential can
be derived in a similar manner as for the $\pi \pi$ and $\pi N$
systems. These ideas have been extensively studied by the Texas--Seattle group
\cite{TEX}. They have obtained an energy dependent two--nucleon potential\footnote{Such explicit energy
dependence of the potential complicates its application to three-- and more--nucleon problems.} 
at NNLO. With altogether 26 free 
parameters\footnote{Some of them
are redundant and can be eliminated from the Lagrangian via Fierz transformation.} 
they were able to achieve a qualitative agreement 
with the NN data.

Alternatively, one can also do the expansion directly on the
level of the scattering amplitude, as it has been proposed by Kaplan 
et al. \cite{Ka98}. In their power counting scheme the leading order 
non--perturbative contribution in the S--channels is given by 
resummation of the iterated non--derivative NN contact interaction. 
All corrections to this result including those from various pion exchanges 
are treated as small perturbations \cite{KSW}.

In what follows we will concentrate on the potential approach
to the few--nucleon scattering. 
In ref.~\cite{egm} (referred to as I from here on)
we constructed the NN and 3N potential based
on the most general chiral effective pion--nucleon Lagrangian using
the method of unitary transformations. For that, we developed a power
counting scheme consistent with this projection formalism. In contrast
to previous results obtained in old--fashioned time--ordered perturbation 
theory, the method employed leads to energy--independent potentials.
To LO, the potential consists of the
undisputed OPE  and two NN contact interactions.
Corrections at NLO stem from the leading chiral TPE and additional contact terms with two derivatives.
At NNLO one has to include the subleading TPE which contains
pion--nucleon interactions with two derivatives (pion mass insertions). 
In contrast to ref.~\cite{TEX}, we take the
novel LECs from systematic studies of
pion--nucleon scattering in
CHPT~\cite{paul}. 
The resulting effective potential is renormalized and applied to two--body  
bound state and scattering problems \cite{egm2} (referred to as II). 

We also consider 3N and 4N systems within the chiral effective field theory \cite{eplast}  (referred to as III).
To the best of our knowledge this is the first
time that $\chi$PT has been practically applied to nuclear systems beyond $A=2$
within the Hamiltonian approach. In this first application we restrict ourselves to the NLO NN potential.
At this order no three--nucleon forces (3NFs) occur\footnote{The 3NF appears first at NNLO and contains
5 undetermined parameters.} \cite{egm}. The NLO results presented here are therefore parameter free and can
serve as a good testing ground for the usefulness of the approach.
Of course, some aspects of the 3N system have already been studied in
nuclear EFT~\cite{Bedaque,Griess}, but not as direct extensions of the
NN system as done here.

Our manuscript is organized as follows. In Section 2 we will introduce the effective NN potential
up to NNLO and give explicit expressions in momentum space. In Section 3 we apply it to the 
NN bound state and scattering problems. In the next section 4 we discuss NLO results for 3N and 4N systems.
Our findings are summarized in Section 5.

\section{THE NN POTENTIAL}

In~I, we derived the two--nucleon potential at next--to--leading order.
To leading order, it consists of the one--pion--exchange potential (OPEP)
and two S--wave 
(non--derivative) contact interactions. The latter are parametrized by
two LECs $C_S$ and $C_T$ \cite{We90},
\beq
\label{LO}
V_{NN}^{LO}=-\biggl(\frac{g_A}{2f_\pi}\biggr)^2 \, \vec{\tau}_1 \cdot
\vec{\tau}_2 \, \frac{\vec{\sigma}_1 \cdot\vec{q}\,\vec{\sigma}_2\cdot\vec{q}}
{q^2 + M_\pi^2} + C_S  + C_T \, \vec{\sigma}_1 \cdot  \vec{\sigma}_2~,
\eeq
where $f_\pi = 92.4\,$MeV
is the pion decay constant, $M_\pi = 138.03\,$MeV the pion
mass and $g_A = 1.26$ the axial--vector coupling. Further, $\vec{q} = \vec{p}~'-\vec{p}$, 
where $\vec{p}$ and $\vec{p}~'$ stand for initial and final nucleon momenta, respectively.
Clearly, the LO potential can only serve as a very crude approximation of the NN force.
\begin{figure}[htb]
\centerline{
\psfig{file=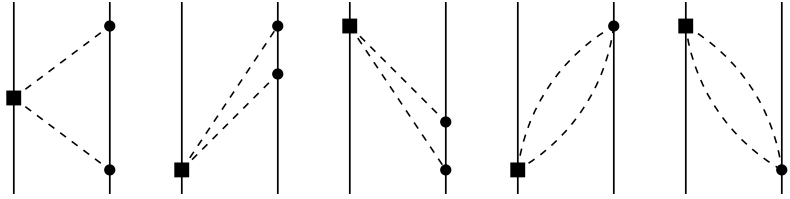,width=4.1125in}}
\vspace{-0.3cm}
\caption[fig19]{\label{fig2}   NNLO corrections to the NN potential.
The filled squares denote $\pi \pi NN$ vertices with two derivatives.}
\vspace{0.0cm}
\end{figure}
At NLO order,
we have three distinct contributions. The first one represents the one--loop
self--energy and vertex corrections to the OPEP and to LO contact interactions.
As discussed in II and \cite{kbw}, such contributions do not lead to form factors and only 
renormalize the appropriate couplings. Secondly, one has seven different NN contact interactions
with two derivatives: 
\beqa
\label{cont}
V_{NN, {\rm cont}}&=&C_1 \, \vec{q}\,^2 + C_2 \, \vec{k}^2 +
( C_3 \, \vec{q}\,^2 + C_4 \, \vec{k}^2 ) ( \vec{\sigma}_1 \cdot \vec{\sigma}_2)
+ iC_5\, \frac{1}{2} \, ( \vec{\sigma}_1 + \vec{\sigma}_2) \cdot ( \vec{q} \times
\vec{k})\no\\
&+& C_6 \, (\vec{q}\cdot \vec{\sigma}_1 )(\vec{q}\cdot \vec{\sigma}_2 ) 
+ C_7 \, (\vec{k}\cdot \vec{\sigma}_1 )(\vec{k}\cdot \vec{\sigma}_2 )~.
\eeqa
Thirdly, there is the genuine TPEP. 
After performing the renormalization it reads:
\beqa\label{TPEP}
V^{\rm TPEP}_{\rm NLO} 
&=& - \frac{ \vec{\tau}_1 \cdot \vec{\tau}_2 }{384 \pi^2 f_\pi^4}\,
L(q) \, \biggl\{4M_\pi^2 (5g_A^4 - 4g_A^2 -1) + q^2(23g_A^4 - 10g_A^2 -1)
+ \frac{48 g_A^4 M_\pi^4}{4 M_\pi^2 + q^2} \biggr\}\no \\
&& - \frac{3 g_A^4}{64 \pi^2 f_\pi^4} \,L(q)  \, \biggl\{
\vec{\sigma}_1 \cdot\vec{q}\,\vec{\sigma}_2\cdot\vec{q} - q^2 \, 
\vec{\sigma}_1 \cdot\vec{\sigma}_2 \biggr\} + P(\vec{k}, \vec{q}\,)~,
\eeqa
with
\beq\label{Lq}
L(q) = \frac{1}{q}\sqrt{4 M_\pi^2 + q^2}\, 
\ln\frac{\sqrt{4 M_\pi^2 + q^2}+q}{2M_\pi}~.
\eeq
Here $P(\vec{k}, \vec{q}\,)$ represents the polynomial part.

At NNLO we have the additional TPEP with one $\pi \pi NN$ vertex with two derivatives as shown in Figure \ref{fig2}.
Three different interactions of this type enter the effective Lagrangian. For the corresponding 
LECs  $c_{1,3,4}$ we take the values obtained from fitting the invariant amplitudes
inside the Mandelstam triangle, i.e. in the unphysical region~\cite{paul}. 
Various NNLO self--energy and vertex correction 
diagrams lead again only to renormalization of the corresponding couplings. Note further that the contact 
interactions are renormalized by an infinite amount in order to absorb all UV divergences of the NNLO TPEP.
The explicit expressions for the renormalized NNLO TPEP can be found in II and in \cite{kbw}.

The effective potential discussed above is only meaningful for 
momenta below a certain scale and therefore needs regularization.
That is done in the following way:
\beq\label{Vreg}
V( \vec{p},\vec{p}~'\,) \to f_\Lambda ( \vec{p}\,) \, V( \vec{p},\vec{p}~'\,) \,
f_\Lambda (\vec{p}~'\,)~,
\eeq
where $f_\Lambda ( \vec{p}\,)$ is a regulator function. 
Here we use two different regulator functions:
\beq
 \label{reg1}
 f_\Lambda^{\rm sharp} ( \vec{p}\,) = \theta (\Lambda^2 -p^2)~, \quad \quad \quad \quad \quad
 f_\Lambda^{\rm expon} ( \vec{p}\,) = \exp(-p^{4} / \Lambda^{4})~.
\eeq

\section{RESULTS FOR THE NN SYSTEM}

Having introduced the effective potential up to NNLO we will now apply it to 
the NN system.

\subsection{Phase shifts}

In order to apply the effective potential to bound state and scattering problems, one first 
needs to determine the unknown couplings related to the NN contact interactions.
At LO one has two such constants ($C_S$, $C_T$), whereas nine constants ($C_S$, $C_T$,
$C_1 \ldots C_7$) occur at NLO and NNLO, see eqs.~(\ref{LO}), (\ref{cont}).
To pin down this couplings we do not
perform global fits as done in ref.~\cite{TEX}. Rather we introduce 
independent new parameters by projections on to appropriate 
partial waves. To leading order,
\begin{figure}[hbt]
\psfrag{1S0}{\hskip -0.5 true cm \small $^1S_0$ [deg]}
\psfrag{3S1}{\hskip -0.5 true cm \small $^3S_1$ [deg]}
\psfrag{1P1}{\hskip -0.5 true cm \small $^1P_1$ [deg]}
\psfrag{1D2}{\hskip -0.5 true cm \small $^1D_2$ [deg]}
\psfrag{3P0}{\hskip -0.5 true cm \small $^3P_0$ [deg]}
\psfrag{3D3}{\hskip -0.5 true cm \small $^3D_3$ [deg]}
\psfrag{3F4}{\hskip -0.5 true cm \small $^3F_4$ [deg]}
\psfrag{3G5}{\hskip -0.5 true cm \small $^3G_5$ [deg]}
\psfrag{E1}{\hskip -0.5 true cm \small $\epsilon_1$ [deg]}
\parbox{5cm}{\centerline{\psfig{file=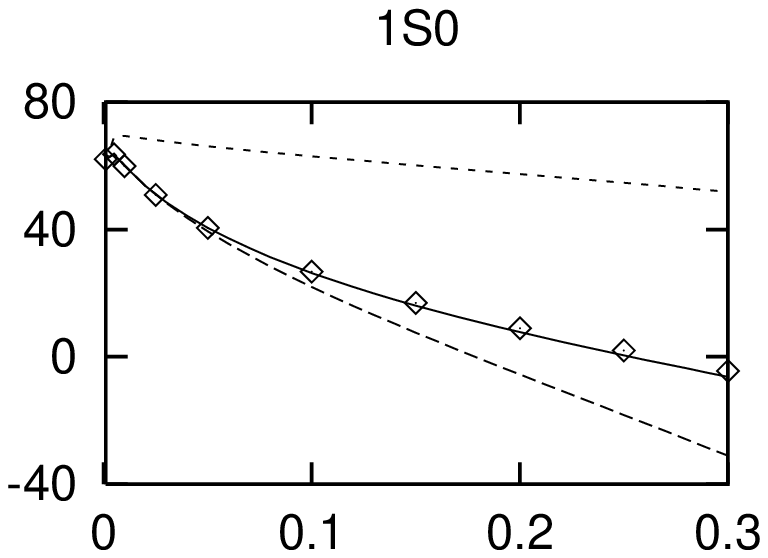,width=6cm}}}
\hfill
\parbox{5cm}{\centerline{\psfig{file=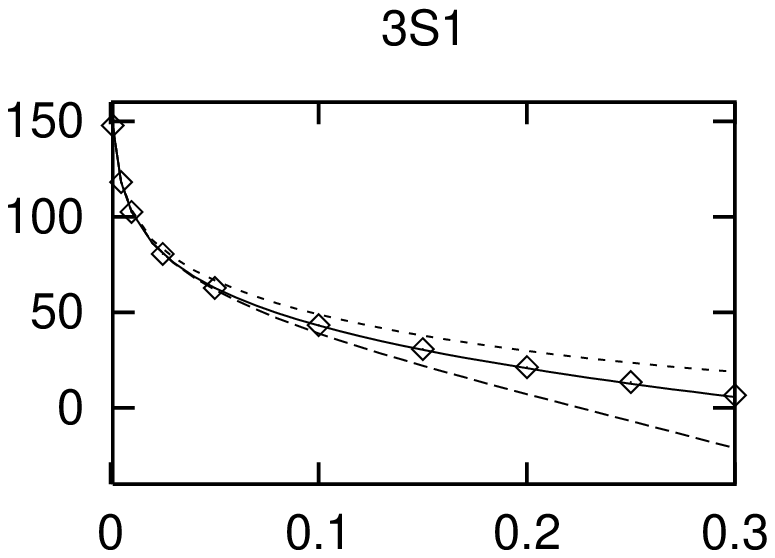,width=6cm}}}
\hfill
\parbox{5cm}{\centerline{\psfig{file=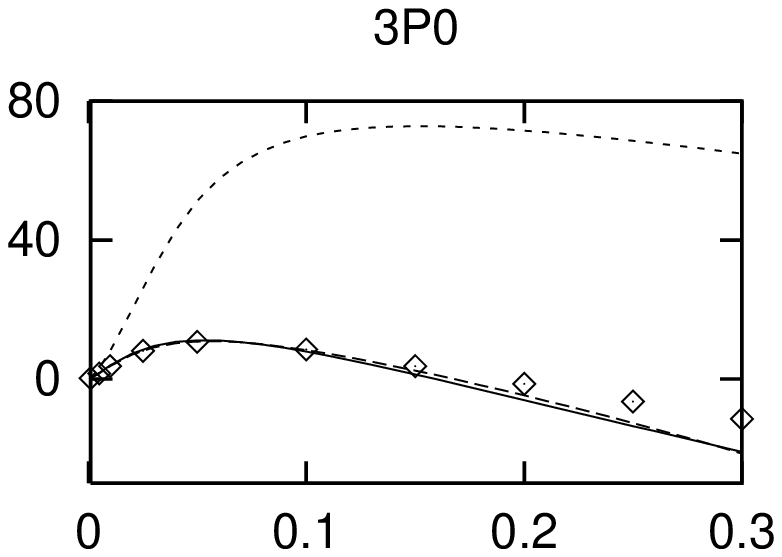,width=6cm}}}
\parbox{5cm}{\centerline{\psfig{file=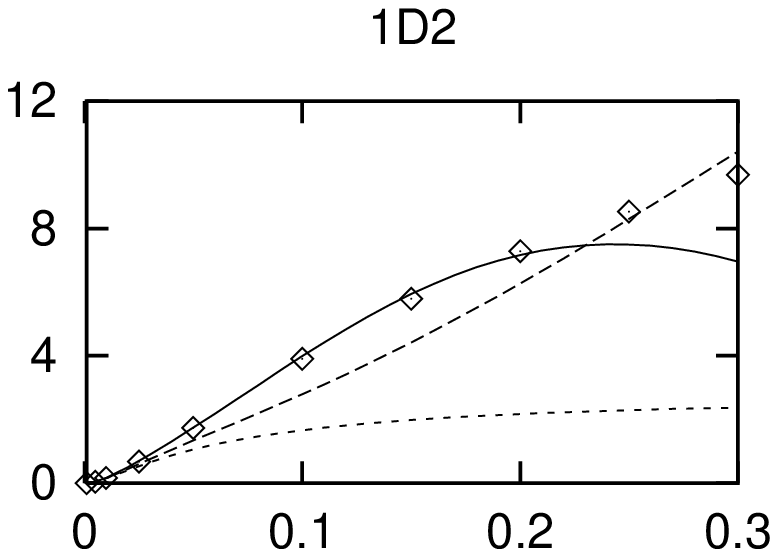,width=6cm}}}
\hfill
\parbox{5cm}{\centerline{\psfig{file=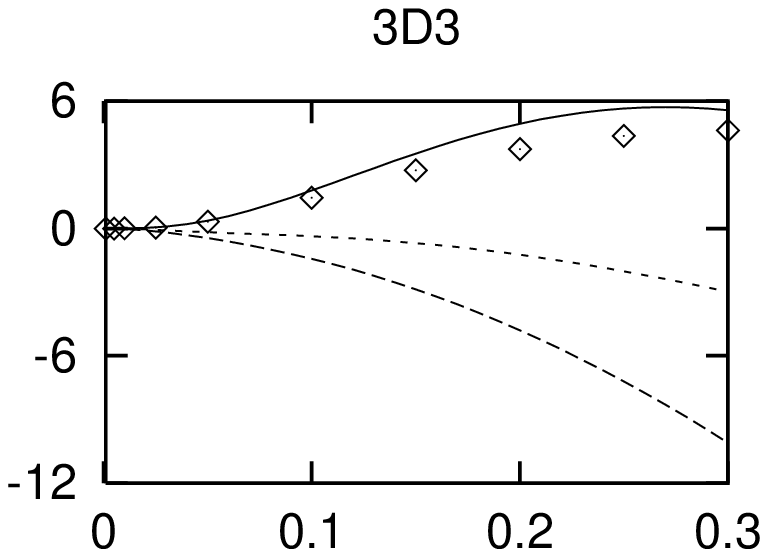,width=6cm}}}
\hfill
\parbox{5cm}{\centerline{\psfig{file=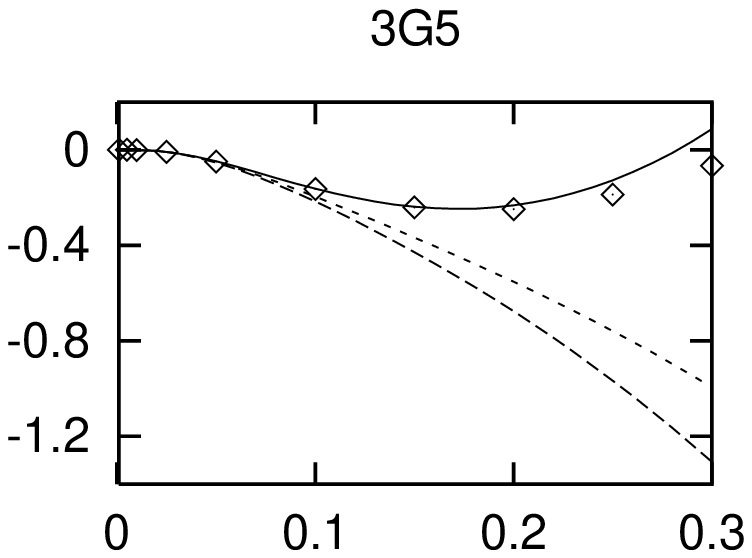,width=6cm}}}
\caption{\label{fig3}
Predictions for various partial wave phase shifts (in degrees) for 
nucleon laboratory energies $E_{\rm lab}$ below 300~MeV (0.3~GeV).
The dotted, dashed and  solid 
curves represent LO, NLO and NNLO results, in order. The 
squares depict the Nijmegen PSA results.}
\end{figure}
the two S--waves are depending on one parameter each. At NLO and NNLO, we have one
additional parameter for $^1S_0$ and $^3S_1$ as well as one parameter in each of the
four P--waves and in $\epsilon_1$. We thus can fit each partial wave separately,
which makes the fitting procedure not only extremely simple but also
unique. All fits have been performed to the phase shifts of the Nijmegen phase shift
analysis (NPSA) \cite{Nij93} for laboratory energies smaller than
(50)~100~MeV at (NLO)~NNLO as described in detail in II.
Having determined the parameters, we can now predict the S-- and P--waves
for energies above 100~MeV. All other partial waves are parameter
free predictions for all energies considered. In Figure \ref{fig3} we show the 
results using the sharp regulator function with $\Lambda =
500$ and $875\,$MeV for LO (NLO) and NNLO, as discussed in II.
At NNLO both S--waves agree well with the data. In the case of $^1S_0$
the OPEP  plus non--derivative contact terms is insufficient to describe the phase shift 
(as it is well--known from
effective range theory and previous studies in EFT approaches).
This also holds true for some P--waves.  
It is known that
TPEP alone is too strong in some of the D-- and F--waves. To the order
we are working, there are no contact interactions so that apart from
varying the cut--off we have no freedom here. Still, for our best global
fit, the D--waves are quite reasonably described.
The result for the $^3D_3$ phase is rather astonishing, since e.g.
in the Bonn potential~\cite{bonn} the correlated two--pion exchange is
very important in the description of  $^3D_3$. Such $\pi\pi$ correlations are only
implicitly present at NNLO. They are hidden in the strengths 
of the LECs $c_1$ and $c_3$\cite{bkmlec}.
From the higher partial waves the $^3G_5$ is the most interesting one. As in the case of $^3D_3$, the OPEP 
alone is not sufficient to describe the phase shifts at energies above 100 MeV, as it could be 
expected for such a high value of angular momentum.
The leading TPEP makes the situation even worse. Only inclusion of the subleading TPEP at NNLO
allows for a very accurate and parameter free\footnote{The cut--off dependence in this channel is rather weak.} 
description of the phase shift.
\begin{table}[htb]
\caption{Scattering lengths, effective range and shape parameters for the $^3S_1$ wave at NLO and 
 NNLO compared to the NPSA.}
\label{tab1}
\newcommand{\m}{\hphantom{$-$}}
\newcommand{\cc}[1]{\multicolumn{1}{c}{#1}}
\renewcommand{\tabcolsep}{1.5pc} 
\renewcommand{\arraystretch}{1.2} 
\begin{center}
\begin{tabular}{lccccc}
    \hline
 & a [fm] & r [fm] & $v_2$ [fm$^3$] & $v_3$ [fm$^5$] & $v_4$ [fm$^7$] \\
    \hline   
$^3S_1$ NLO         &   5.434 & 1.711 & 0.075 & 0.77  & $-$4.2  \\
$^3S_1$ NNLO        &   5.424 & 1.741 & 0.046 & 0.67  & $-$3.9  \\
$^3S_1$ NPSA        &   5.420 & 1.753 & 0.040 & 0.67  & $-$4.0  \\
    \hline
  \end{tabular}
\end{center}
\end{table}

We have also calculated the values of various effective range and shape parameters for $^3S_1$.
In Table \ref{tab1} we compare our NLO and NNLO predictions for these quantities with the results 
from the Nijmegen PSA.
As expected, in all cases the NNLO results  
go closer to the data than the NLO ones. Note further that in both cases one 
has the same number of free parameters. Therefore we consider the improvement as an indication 
of a good convergence of our expansion.

\subsection{Deuteron}
We now turn to the bound state properties. At NNLO (NLO), we consider an exponential
regulator with $\Lambda =1.05\,(0.60)\,$GeV, which reproduces the deuteron
binding energy $E_{\rm d}=-2.224575(9)$ MeV within an accuracy of about one third of a permille
(2.5 percent)\footnote{Note
that the deuteron binding energy is {\it not} used to fit the free
parameters in the potential.}: $E_{\rm d}=-2.2238$ (-2.1650) MeV.
The asymptotic $D/S$ ratio,
called $\eta$, and the strength of the asymptotic wave function, $A_S$, as well as 
the root--mean--square matter radius $r_d$ are well
described at NNLO (NLO): $\eta=0.0245$ (0.0248), 
$A_{\rm S}=0.884$ (0.866) fm$^{-1/2}$ and 
$r_{\rm d}=1.967$ (1.975) fm compared to experimentally
measured values $\eta=0.0256(4)$, $A_{\rm S}=0.8846(16)$ fm$^{-1/2}$ and 
$r_{\rm d} =1.9671(6)$ fm, respectively.
The D--state probability, which is not an observable, seems to be 
most sensitive to small variations in the cut--off. 
We observe a small deviation in the 
values for the quadrupole moment, $Q_{\rm d}=0.262$ (0.266) fm$^2$
compared to the experimental value $Q_{\rm d}=0.2859(3)$  fm$^2$,
which is also typical for conventional potentials and possibly arises from the missing exchange current operators.


\section{RESULTS FOR THE 3N AND 4N SYSTEMS}

\subsection{Bound states}
In this section we present the binding energies for 3N and 4N systems calculated from our
NLO potential. As already pointed out above no 3NF appear at this order. 
\begin{table}[htb]
\caption{Theoretical $^3$H and $^4$He binding energies for
     different cut-offs $\Lambda$  
     compared to CD-Bonn predictions and to the experimental $^3$H binding energy and the Coulomb
     corrected $^4$He binding energy.}
\label{tab3}
\newcommand{\m}{\hphantom{$-$}}
\newcommand{\cc}[1]{\multicolumn{1}{c}{#1}}
\renewcommand{\tabcolsep}{0.75pc} 
\renewcommand{\arraystretch}{1.2} 
    \begin{center}
    \begin{tabular}[t]{ccccccc}
\hline
& NLO 540 & NLO 560 & NLO 580 & NLO 600 & CD Bonn & Exp. \\ \hline 
 $E_{^3 \rm{H}}$ [MeV] &  -8.284 & -8.091 & -7.847 & -7.564 & -8.012 & -8.48 \\
$E_{^4\rm{He}}$ [MeV] & -28.03 & -26.91 & -25.55 & -23.96 & -27.05 & -29.00 \\ \hline
     \end{tabular}
     \end{center}
\end{table}
The NLO results shown here are therefore parameter free predictions and can
serve as a good testing ground for the usefulness of the approach.
In order to investigate the cut-off dependence of 3N and 4N observables we
have generated several NN potentials corresponding to different exponential cut-offs
between $\Lambda$ =540 and 600 MeV. They were all fitted to the $^1S_0$,
$^3S_1$-$^3D_1$, $^1P_1$ and $^3P_{0,1,2}$ NN phase shifts 
up to $E_{\rm lab} = 100\,$MeV, see III for more details.
We find for the fully
converged solutions of the corresponding Faddeev-Yakubovsky equations 
the 3N and 4N binding energies as given in Table \ref{tab3}.
The ranges are compatible with what is found using realistic
potentials. Note that the
NN forces were included up to the total 
NN angular momentum of $j_{max}$=6.


\subsection{{\it nd} scattering}
\begin{figure}[hbt]
\vspace{-0.7cm}
\psfrag{xxx}{\raisebox{-0.17cm}{\hskip -0.3 true cm  $\theta$ [deg]}}
\psfrag{dsig}{\raisebox{-2.0cm}{\hskip -1.6 true cm  $d \sigma$}}
\psfrag{ay}{\raisebox{-0.9cm}{\hskip -1.6 true cm $A_y$}}
\psfrag{t20}{\raisebox{-0.9cm}{\hskip -1.4 true cm  $T_{20}$}}
\parbox{5cm}{\centerline{\psfig{file=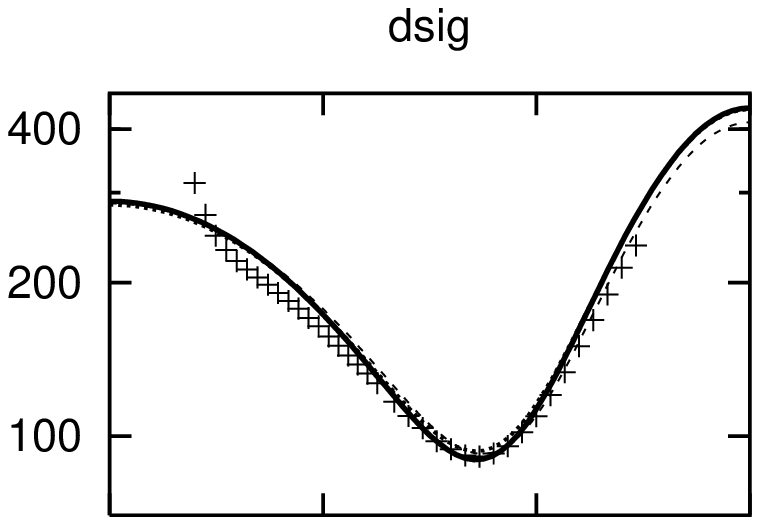,width=6cm,height=4.3cm}}}
\hfill
\parbox{5cm}{\centerline{\hskip -0.2 true cm \psfig{file=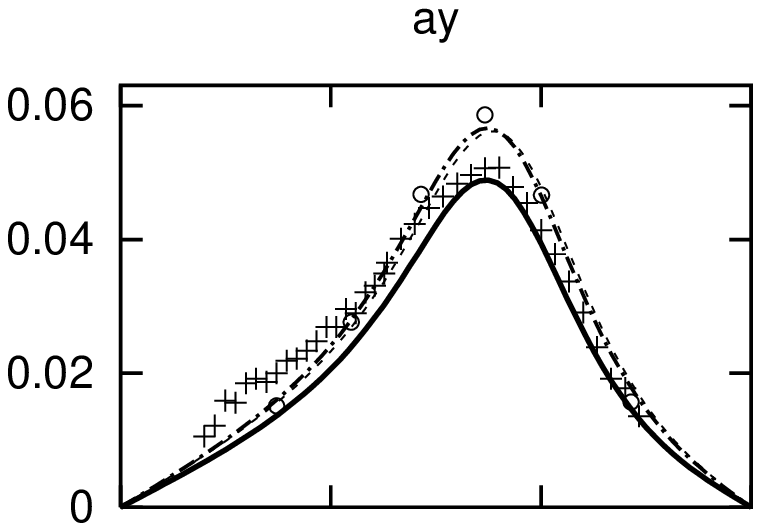,width=6.2cm,height=4.3cm}}}
\hfill
\parbox{5cm}{\centerline{\hskip -0.2 true cm \psfig{file=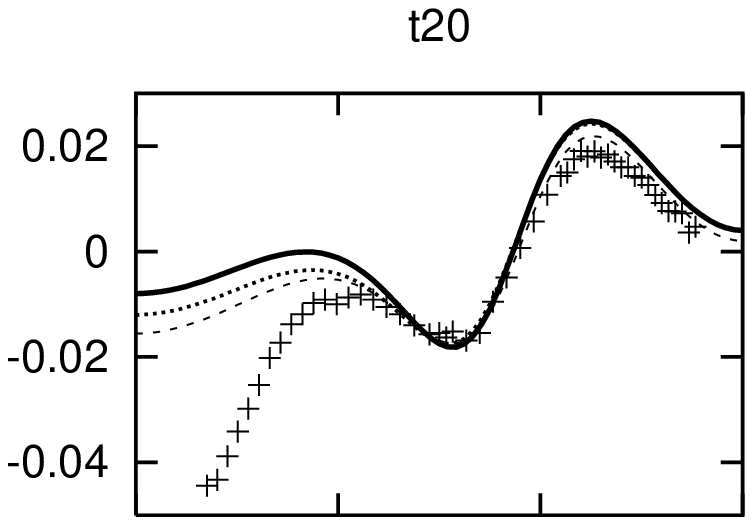,width=6.2cm,height=4.3cm}}} 
{\vskip -1 true cm
\parbox{5cm}{\centerline{\psfig{file=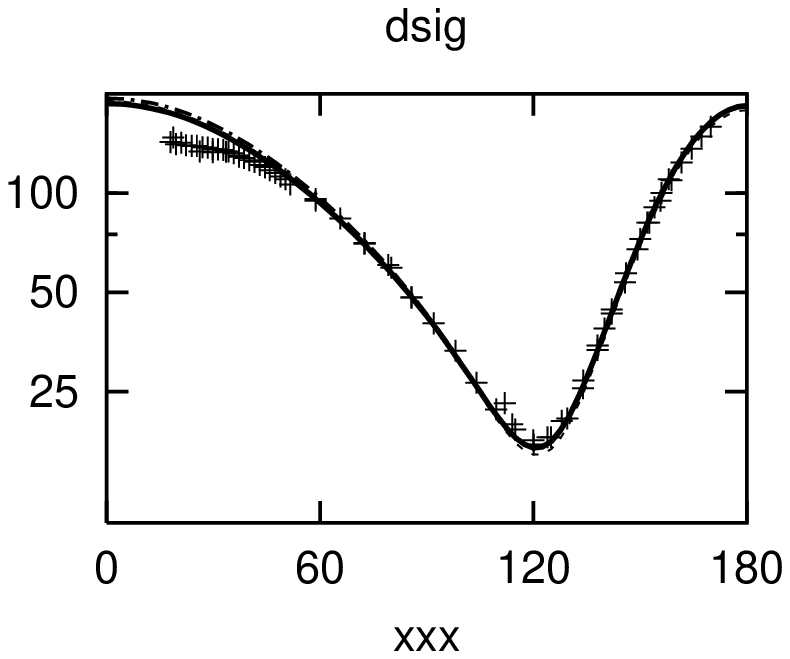,width=6cm,height=5cm}}}
\hfill
\parbox{5cm}{\centerline{\psfig{file=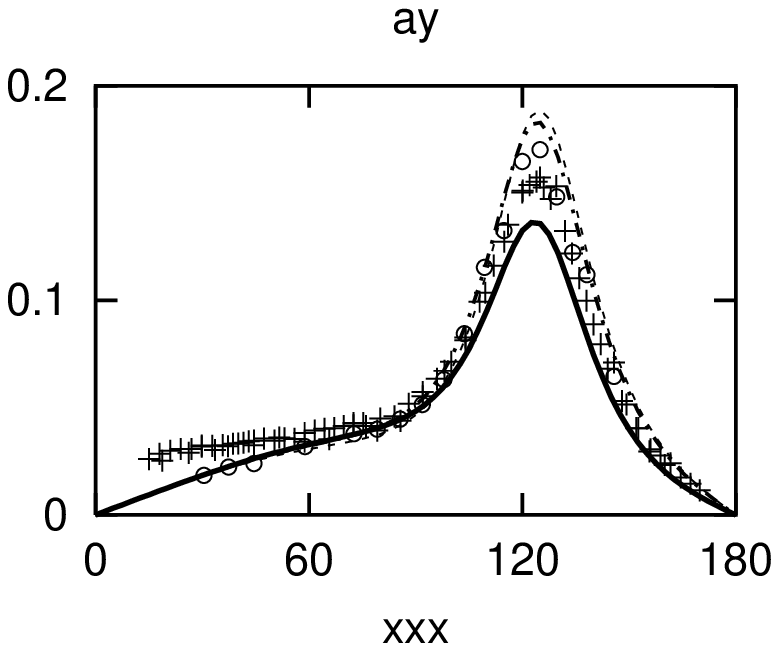,width=6cm,height=5cm}}}
\hfill
\parbox{5cm}{\centerline{\psfig{file=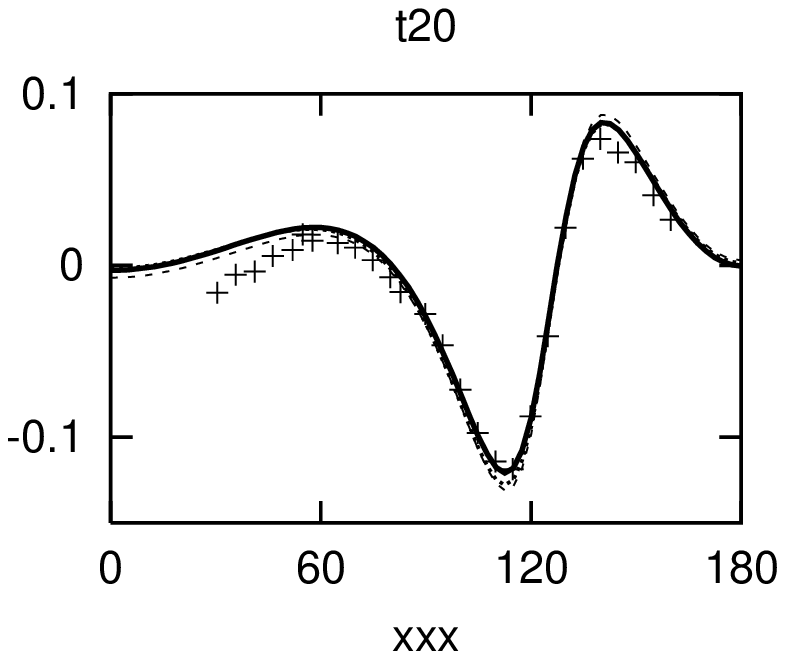,width=6cm,height=5cm}}}}
\caption[fig19]{\label{fig4} {\it nd} elastic scattering observables at $E_n=3$MeV (upper row)
     and $E_n=10$MeV (lower
    row) for the chiral forces ($\Lambda = 540$~MeV/c,
    dotted curve; $\Lambda = 600$~MeV/c, dashed curve), and CD-Bonn (thick solid curve). 
    {\it pd} data are ($\scriptstyle +$) \protect\cite{Sagara}. 
    {\it nd} data are ($\circ$) \protect\cite{McAninch}.}
\end{figure}

Also, 3N scattering can be solved rigorously nowadays \cite{Gloeckle} 
and we show in Figs.~\ref{fig4} some elastic observables as a particular example, namely
 the angular
distributions, the nucleon analyzing power $A_y$ and the tensor analyzing power $T_{20}$.
Except for $A_y$ there are no {\it nd} data for those energies.
The discrepancies between data and theory for $T_{2k}$ and for the differential cross section
can be traced back to the {\it pp} Coulomb force \cite{Kievsky}. 
Thus except for $A_y$ the agreement of CD-Bonn with the data is rather
good, which is a well known fact and is just given for the sake of 
orientation. Our parameter free predictions for $d \sigma$ and $T_{2k}$ agree rather well with the CD-Bonn
result and thus with 
the data. We consider this to be an important result, demonstrating that the 
chiral NN forces are very well suited to also describe low-energy 3N scattering observables
rather quantitatively. On top of that, 
surprisingly for us, the chiral force predictions are now
significantly higher in the 
maxima of $A_y$ than for CD-Bonn and break the long standing situation, that
all standard realistic NN 
forces up to now underpredict the maxima by about 30~\%. That
used to be called 
the $A_y$--puzzle\cite{Gloeckle}. We are now in fact rather close to the 
experimental {\it nd} values. Note that on a 2N level
the chiral potential predictions for the analyzing power agree well with the predictions based on the 
Nijmegen PSA \cite{eplast}. Our predictions for various break up observables agree also very well
with the ones obtained from CD-Bonn, as it is demonstrated in \cite{eplast}.

\section{CONCLUSIONS}

Our findings show that the scheme originally proposed by
Weinberg works quantitatively and much better than it was
expected. It extends the successfull applications of effective field
theory (chiral perturbation
theory) in the pion and pion--nucleon sectors to systems with more than
one nucleon. Most of the {\it np} partial waves are well described at NNLO.
The NNLO TPEP turnes out to be  too strong in the triplet F--waves. This is expected to be 
cured at N$^3$LO.

The very first results using chiral NN forces in 3N and 4N systems
are very promising and show that these 
effective chiral forces are very well suited to describe also  low
energy properties of 
nuclear systems beyond $A=2$. They agree rather well with standard nuclear force predictions as
exemplified with CD-Bonn 
and most importantly they break the stagnation in the $A_y$--puzzle.
It will be very interesting to perform the
next step and use the 
NNLO NN forces, which would also require inclusion of the leading 3NF.


\begin{thebibliography}{9}
\bibitem{We79} S.~Weinberg, Physica A96 (1979) 327. 
\bibitem{We90} S.~Weinberg, Phys. Lett. B251 (1990) 288, Nucl. Phys. B363 (1991) 3. 
\bibitem{TEX} C.~Ord\'o\~nez and U.~van Kolck, Phys. Lett. B291 (1992) 459, 
Phys. Rev. Lett. 72 (1994) 1982, Phys. Rev. C53 (1996) 2086.
\bibitem{Ka98} D.B.~Kaplan, M.J.~Savage and M.B.~Wise, Phys. Lett. B424 (1998) 390. 
\bibitem{KSW} D.B.~Kaplan, M.J.~Savage and M.B.~Wise, Nucl. Phys. B534 (1998) 329,
 Phys. Rev. C59 (1999) 617; M.J.~Savage and R.P.~Springer, Nucl. Phys. A644 (1998) 235;
E.~Epelbaum, Ulf-G.~Mei\3ner, Phys. Lett. B461 (1999) 287; S.~Fleming, T.~Mehen, and I.W.~Stewart, 
Nucl. Phys. A677 (2000) 313.
\bibitem{egm}E. Epelbaoum, W. Gl\"ockle and Ulf-G. Mei{\ss}ner,
  Nucl. Phys. A637 (1998) 107.
\bibitem{paul} P. B\"uttiker and Ulf-G. Mei{\ss}ner, Nucl. Phys. A668 (2000) 97.
\bibitem{egm2}E. Epelbaoum, W. Gl\"ockle and Ulf-G. Mei{\ss}ner,
  Nucl. Phys. A671 (2000) 295.
\bibitem{eplast} E.~Epelbaum, {\it et al.},
{\tt nucl-th/0007057}.
\bibitem{Bedaque}P.F.~Bedaque, H.--W.~Hammer, and U.~van Kolck, {\it et al.},
Phys. Rev. C58 (1998) R641;  Phys. Rev. Lett.  82
(1999) 463.
\bibitem{Griess}F.~Gabbiani, P.F.~Bedaque, and H.~Griesshammer,
Nucl. Phys. A675 (2000) 601.
\bibitem{kbw} N.~Kaiser, R.~Brockmann and W.~Weise, 
              Nucl. Phys. A625 (1997) 758.
\bibitem{Nij93} V.G.J.~Stoks, {\it et al.},
Phys. Rev. C49 (1994) 2950.
\bibitem{bonn}R. Machleidt, K. Holinde and Ch. Elster,
Phys. Rep. 149 (1987) 1.
\bibitem{bkmlec}V.~Bernard, N.~Kaiser and Ulf-G.~Mei{\ss}ner,
                Nucl. Phys. A615  (1997) 483.
\bibitem{Gloeckle}
 W.~Gl\"ockle {\it et al.}, 
Phys. Rep.  274 (1996) 107.
\bibitem{Sagara} 
K.~Sagara {\it et al.},
 Phys. Rev.  C50 (1994) 576; 
S.~Shimizu {\it et al.},
 Phys. Rev. C52 (1995) 1193;
G.~Rauprich {\it et al.}, Few-Body Systems 5 (1988) 67;
F.~Sperisen  {\it et al.}, Nucl. Phys.  A422 (1984) 81.
\bibitem{McAninch} 
J.E.~McAninch {\it et al.},
 Phys. Lett.  B307 (1993) 13;
C. R.~Howell  {\it et al.},
 Few Body Systems   2 (1987) 19.
\bibitem{Kievsky}
 A. Kievsky, Phys. Rev. C60 (1999) 034001.
\end{thebibliography}
\end{document}